\documentclass[aps,prb,preprintnumbers,showkeys,twocolumn,amsmath,amssymb,longbibliography,floatfix]{revtex4-2}%

\usepackage[pdftex,colorlinks,hyperindex,plainpages=false,bookmarksopen,bookmarksnumbered,pdfusetitle,allcolors={blue}]{hyperref}
\usepackage{graphicx}
\usepackage{dcolumn}
\usepackage{bm}
\usepackage[capitalize]{cleveref}
\usepackage{svg}
\usepackage{color}
\usepackage{amsmath}
\usepackage{epstopdf}
\usepackage{fancybox}
\usepackage{amsfonts}
\usepackage{amssymb}%
\usepackage[toc,page]{appendix}
\usepackage{adjustbox}

\usepackage[utf8]{inputenc}
\usepackage[T1]{fontenc}

\renewcommand{\cite}[1]{{[}\onlinecite{#1}{]}}

\newcommand{\be}{\begin{equation}}
\newcommand{\e}{\end{equation}}
\newcommand{\beml}{\begin{subequations}}
\newcommand{\eml}{\end{subequations}}
\newcommand{\beq}{\begin{eqnarray}}
\newcommand{\eq}{\end{eqnarray}}
\newcommand{\ba}{\begin{array}}
\newcommand{\ea}{\end{array}}
\newcommand{\bpm}{\begin{pmatrix}}
\newcommand{\epm}{\end{pmatrix}}
\newcommand{\bc}{\begin{cases}}
\newcommand{\ec}{\end{cases}}

\DeclareMathOperator{\tr}{Tr}

\begin{document}
\title{Light-induced Dzyaloshinskii-Moriya interactions in antiferromagnetic metals}

\author{ Sander Ø. Hanslin}
\affiliation{Center for Quantum Spintronics, Department of Physics, Norwegian University of Science and Technology, NO-7491 Trondheim, Norway}
\author{Alireza Qaiumzadeh}
\affiliation{Center for Quantum Spintronics, Department of Physics, Norwegian University of Science and Technology, NO-7491 Trondheim, Norway}

\date{\today}

\begin{abstract}
The Dzyaloshinskii-Moriya (DM) interaction plays an essential role in novel topological spintronics, and the ability to control this chiral interaction is of key importance. Developing a general microscopic framework to compute DM interactions using imaginary-time Green's function formalism, we theoretically show that the ac electric field component of a laser pulse induces {\it{nonequilibrium}} static DM interactions in an antiferromagnetic system in the presence of relativistic spin-orbit coupling. These induced DM interactions might even be anisotropic depending on the direction of magnetic moments and the laser pulse polarization. We further show that intense polarized laser pulses can in principle generate both classes of DM interactions, i.e., {\it{bulk}}-type and {\it{interfacial}}-type, in a magnetic system even though the crystal symmetry prohibits one of them in equilibrium. Our results reveal another aspect of rich behavior of periodically driven spin systems and out of equilibrium magnetic systems.
\end{abstract}

\maketitle

\section{Introduction}
The Dzyaloshinskii-Moriya (DM) interaction is an antisymmetric exchange interaction arising in magnetic systems with broken spatial inversion symmetry. This chiral interaction is responsible for many exotic states and the stabilization of topological solitons in both ferromagnetic (FM) and antiferromagnetic (AFM) systems, such as helimagnets \cite{Meier}, chiral domain walls \cite{Legrand2018,PhysRevB.78.140403}, skyrmions \cite{Muhlbauer915,Fert2017,PhysRevB.99.054423,Antiskyrmion}, hopfions \cite{PhysRevLett.123.147203}, topological magnons \cite{topomagnon}, and nonreciprocal dynamics of magnetic excitations or magnons \cite{PhysRevLett.104.137203,PhysRevLett.114.087203}. In AFM materials, DM interactions also break the degeneracy of two magnon modes with opposite helicities \cite{PhysRevLett.119.047201,Nowak,PhysRevB.97.020402}, making AFM materials promising for magnonic devices.

The competition between the DM interaction and the Heisenberg exchange defines the degree of deviation from collinearity of the spin structures and also governs the stability of topological spin textures. A sudden change in this ratio can lead to excitation of ultrafast magnons in the system \cite{Mikhaylovskiy2015}. Also dynamical control of topological spin textures is of great interest \cite{PhysRevLett.118.157201}.
Hence, finding an efficient method of controlling DM interactions is one of the current technological interests and challenges in novel nanoscale spintronics. From a fundamental point of view, the precise and systematic calculation of DM interactions in different magnetic structures as well as the formulation of new methods for inducing equilibrium and nonequilibrium DM interactions in magnetic materials is an open question, and there is an urgent demand for new proposals.

Based on the symmetry group analysis of crystals, Dzyaloshinskii proposed that an antisymmetric exchange interaction is allowed in noncentrosymmetric magnetic materials \cite{DZYALOSHINSKY1958241}. The microscopic origin of the DM interaction was formulated by Moriya using the Anderson superexchange interaction mechanism in the presence of spin-orbit coupling (SOC) \cite{PhysRev.120.91}. Moriya found that the DM interaction is linearly proportional to the SOC strength. Nevertheless, this formalism is not quite suitable for quantitatively computing the DM interactions of real structures. Recent developments in spintronics necessitate more convenient machinery and precise approaches suitable for realistic systems with complicated band structures, e.g., in conjunction with {\it{ab initio}} calculations.
In recent years, there have been several new proposals for computing equilibrium DM interactions in insulating and metallic magnetic systems. Katsnelson {\it{et al.}} developed a method for calculating DM interaction using the exact perturbation expansion of the total energy of an AFM system in the canting angle \cite{PhysRevB.82.100403}.
In another approach, the origin of DM interactions is attributed to a Doppler shift due to an intrinsic spin current induced by SOC \cite{PhysRevLett.116.247201}.
Several different approaches based on the Berry phase formalism, spin-spin correlation functions and the magnetic force theorem have been developed recently to compute the DM interactions of realistic band structures in both FM and AFM systems in equilibrium \cite{PhysRevLett.120.197202,1804.03739,PhysRevB.101.161403,Freimuth_2014,Koretsune2015,PhysRevB.79.045209,PhysRevB.68.104436}.
Equilibrium DM interaction might be enhanced and even induced in magnetic thin films by breaking the inversion symmetry via, e.g., sandwiching a magnetic slab between two different layers, using proximity effects in heavy-metal (HM) and magnetic film heterostructures \cite{Jadaun,PhysRevLett.118.147201,Chshiev,PhysRevLett.111.216601, PhysRevB.102.224414}, or applying electric gate voltages and lattice strains \cite{PhysRevLett.106.247203, PhysRevB.102.134422,PhysRevMaterials.4.094004}.

Nonequilibrium DM interactions have been a subject of very recent studies in spintronics and quantum magnetism. DM interactions in FM metals can be modified by charge currents and electrical voltages \cite{PhysRevB.102.245411,Takeuchi_2019}. On the other hand, it has recently been proposed that DM interactions can be dramatically tuned in the presence of nonresonant laser pulses via a direct coupling between the electric field component of light and spins \cite{Mikhaylovskiy2015,PhysRevLett.118.157201,PhysRevLett.116.125301,losada,Yudin2017}. However, the situation in metallic AFM systems \cite{metallicAFM} is unclear.

In this paper, we study the DM interaction induced in a metallic AFM system with finite SOC under a periodically driven electric field such as that arising from an intense laser pulse. Using the path integral approach and the imaginary-time Matsubara Green's function formalism \cite{Rammer}, we develop a general formalism based on the computation of spin correlation functions. Then, we compute the induced DM interaction and present analytical results for a toy model of a two-dimensional (2D) Rashba AFM system. Our formalism is quite general and can possibly be implemented in first-principles codes to compute induced DM interactions in more complicated conventional AFM systems as well as recently discovered topological and Weyl AFM materials \cite{AFMTI,2018NatPh..14..242S,mejkal_2017}.

The rest of the paper is organized as follows. In Sec. II, we find a relation for the DM interaction tensor in terms of correlation functions. In Sec. III, we introduce the path integral formalism to determine the effective bosonic action for a generic magnetic system. In Sec. IV, we introduce our model Hamiltonian and perturbing potentials. In Sec. V, we review the results for the equilibrium DM interaction induced by the Rashba SOC in an AFM system. In Sec. VI, we derive the DM interaction induced by an ac electric field using 4th-order perturbation theory. We present our conclusions in Sec. VII.

\section{DM interactions in terms of correlation functions}
Phenomenologically, the free energy density of an inhomogeneous DM interaction can be written as \cite{Landau,PhysRevB.101.161403,Bogdanov,PhysRevLett.119.127203}
\begin{align}
\label{microDMI}
E_{\mathrm{DM}}=D_{abc} n^a\partial_b n^c,
\end{align}
where the Einstein summation convention over repeated indices is employed and $a,b,c\in\{x,y,z\}$.
In the above equation, $\bm{n}$ is the order parameter vector field, e.g., the staggered or N{\'e}el vector in AFM systems and the total magnetization vector in FM systems, and $D$ is the DM interaction tensor. As already mentioned, it is possible to find the nonzero elements of the DM tensor based on the magnetic symmetries of crystals. In this section, we want to find a general formula to compute the DM interaction tensor in terms of spin correlation functions \cite{PhysRevLett.120.197202}.

From the definition of the partition function $\mathcal{Z}$ in quantum statistical mechanics, we know that the action $\mathcal{S}$ is related to the total energy of the system $E$,
\begin{align}
\label{ZvsE}
\mathcal{Z}=\int \mathcal{D}\bm{n} e^{-\mathcal{S}_{\mathrm{eff}}[\bm{n}]/\hbar}=\int \mathcal{D}\bm{n} e^{-\beta E[\bm{n}]},
\end{align}
where the integral is taken over a bosonic field, which in our case is the order parameter of the system in question, and $\mathcal{S}_{\mathrm{eff}}$ is the effective bosonic action obtained after summation over the fermionic degrees of freedom of the original total action. The total action is the sum of the fermionic action, bosonic action, and a term describing the interaction between these fermions and bosons.
Using the above relation, one can determine the total magnetic energy of a magnetic system. In this approach, the spin-spin interactions, appearing in the effective micromagnetic energy, are mediated by itinerant electrons and thus have a Ruderman–Kittel–Kasuya–Yosida (RKKY)-like coupling origin \cite{PhysRevLett.44.1538}.
Note that the total micromagnetic energy of the system is the sum of this RKKY-like magnetic energy and the free energy of the localized spins presented in the bosonic spin Hamiltonian.

In general, the effective action can be written in terms of the correlation functions $\mathcal{C}$. Up to the second order deviation of the order parameter from equilibrium $\delta \bm{n}$, we have $\beta \mathcal{S}_{\mathrm{eff}}=\delta n^a \mathcal{C}^{ab} \delta n^b$.
From Eq. (\ref{ZvsE}), we find that $(\beta/\hbar)\delta \mathcal{S}_{\mathrm{eff}}/\delta n^a=\delta E/\delta n^a$. The functional derivatives of the effective action and micromagnetic energy are $\beta\delta \mathcal{S}_{\mathrm{eff}}/\delta n^a= \mathcal{C}^{ab} \delta n^b$ and $\delta E/\delta n^a= D_{abc}\partial_b \delta n^c$, respectively. Expanding the correlation function up to the first order of the wavevector $\bm{q}$, $\mathcal{C}_{\bm{q}}^{ab} \approx -i\hbar(\partial \mathcal{C}_{\bm{q}=0}^{ab}/\partial q_c)\partial_c$, we can determine the DM tensor,
\begin{align}
\label{DMformula}
D_{abc}=-i \frac{\partial \mathcal{C}_{\bm{q}=0}^{ac}}{\partial q_b}.
\end{align}
Thus, different components of the DM interaction tensor are related to the derivative of the spin correlation function with respect to the wave vector.

In the next sections, we compute the appropriate spin correlation functions to find the {\it{equilibrium}} and {\it{light-induced}} DM interactions.

\section{Path integral formalism and effective bosonic action}
In this section, we use thermal quantum field theory and the path integral approach to find the effective bosonic action of a generic magnetic metal in the presence of perturbing potentials. In this approach, we sum over the fermionic degrees of freedom of itinerant electrons, and thus, the resulting effective DM interaction has an RKKY-like origin in which two localized magnetic moments indirectly interact via itinerant electrons.

The total Hamiltonian of a magnetic system is the sum of an unperturbed Hamiltonian and an external perturbing potential $V$ coupled to fermions and can in general be spatially and temporally dependent.
The unperturbed Hamiltonian consists of a fermionic part describing the dynamics of itinerant free electrons and SOC, a bosonic part $\mathcal{H}_{B}$ consisting of the different possible magnetic interactions between localized spins such as the Heisenberg exchange interactions, magnetic anisotropies, and intrinsic DM interactions, with a spin order parameter $\bm{n}$; and an interaction term between the fermions and bosons.

It is more convenient to collect all terms of the total Hamiltonian that have fermionic operators, i.e., the fermionic term, the interaction term, and the perturbing term, in a new total fermionic Hamiltonian $\mathcal{H}_{F}$,
\begin{align}
\label{hamiltonian2}
\mathcal{H}_{F}[\bm{n}]&=\mathcal{H}_{0}[\bm{n}]+V_{\bm{r},\tau}. 
\end{align}
Thus, the total Hamiltonian is $\mathcal{H}_{tot}=\mathcal{H}_{F}[\bm{n}]+\mathcal{H}_{B}[\bm{n}]$.
The total partition function is given by
\begin{align}
\mathcal{Z}&=\int \mathcal{D}\Phi^* \mathcal{D}\Phi \mathcal{D}\bm{n} e^{-(\mathcal{S}_B+\mathcal{S}_F)/\hbar}, \label{partition1}
\end{align}
where $\mathcal{S}_B$ and $\mathcal{S}_F$ are the bosonic and fermionic parts of the action, respectively, and $\Phi$ is the Grassmannian coherent-state spinor.
The fermionic part of the action $\mathcal{S}_F$ related to the fermionic Hamiltonian (\ref{hamiltonian2}) is
\begin{align}
\label{action1}
\mathcal{S}_F&=\int_0^{\hbar\beta}d\tau \int d{\bm{r}} \Phi^*_{\bm{r},\tau} \big(\hbar\partial_\tau + \mathcal{H}_F[\bm{n}]\big) \Phi_{\bm{r},\tau},
\end{align}
where $\beta$ is the thermodynamic beta, $\tau$ is the imaginary time, and $\hbar$ is the reduced Planck's constant.
$\mathcal{S}_B$ similarly describes the dynamics of the localized spins or magnons through the bosonic Hamiltonian $\mathcal{H}_{B}$. In our formalism, we consider a system with strong magnetic anisotropy in the low temperature limit, at which we can ignore both quantum and thermal fluctuations. In this regime, $\mathcal{S}_B$ is integrated out from the partition function.

The total imaginary-time Green's function related to the fermionic Hamiltonian (\ref{hamiltonian2}) is defined as,
\begin{align}
\hbar\mathcal{G}^{-1}_{\bm{r},\tau,\bm{r}',\tau'} =-\left(\hbar\partial_\tau + \mathcal{H}_F\right)\delta(\tau-\tau')\delta(\bm{r}-\bm{r}').
\end{align}
Hence, we can rewrite the fermionic action as
\begin{align}
\label{action2}
\mathcal{S}_F&=\iint_0^{\hbar\beta}d\tau d\tau' \iint d{\bm{r}} d{\bm{r}}' \Phi^*_{\bm{r},\tau} \big(-\hbar\mathcal{G}^{-1}_{\bm{r},\tau,\bm{r}',\tau'}\big) \Phi_{\bm{r}',\tau'}.
\end{align}

To find an effective bosonic action, we integrate all fermionic degrees of freedom in the partition function (\ref{partition1}) using the Gaussian integral technique and Jacobi's formula. Hence, we find $\mathcal{Z}=\int \mathcal{D}\bm{n} e^{-\mathcal{S}_{\mathrm{eff}}[\bm{n}]/\hbar}$, where the effective action can be expressed by
\begin{align}
\label{action3}
\mathcal{S}_{\mathrm{eff}}=-\hbar\iint_0^{\hbar\beta}d\tau d\tau' \iint d{\bm{r}}d{\bm{r}'} \tr\left[\ln \big(-\mathcal{G}^{-1}_{\bm{r},\tau,\bm{r}',\tau'}\big)\right].
\end{align}
The Green's function in this effective action (\ref{action3}) is the total exact Green's function related to the Hamiltonian (\ref{hamiltonian2}) and can be calculated using perturbation theory.

In general, the total perturbation to the equilibrium Hamiltonian can be a linear sum of $l$ different perturbing potentials,
\begin{align}
\label{}
V_{\bm{r},\tau}=\sum_{i=1}^l V^i_{\bm{r},\tau}.
\end{align}
Using the Dyson relation, we find the total Green's function in terms of noninteracting Green's functions $\mathcal{G}^0$,
\begin{align}
\label{Dyson}
\mathcal{G}^{-1}=(\mathcal{G}^0)^{-1}-V.
\end{align}
Inserting Eq. (\ref{Dyson}) into the effective action (\ref{action3}) and expanding the action in terms of the total perturbing potential, in the Fourier space, we can formally write the total effective action as,
\begin{align}
\label{action4}
&\mathcal{S}_{\mathrm{eff}}=\mathcal{S}^{(\mathrm{0})}+\mathcal{S}^{(1)}+\sum_{n=2}^\infty \mathcal{S}^{(n)},\\
&\mathcal{S}^{(n\geq2)}=\sum_{p=1}^{l} \mathcal{S}_{\kappa_p}^{(n)}, 
\label{action5}
\end{align}
where $\mathcal{S}^{(\mathrm{0})}$ and $\mathcal{S}^{(1)}$ are the noninteracting and first-order perturbed actions, respectively. $\mathcal{S}^{(n)}$ with $n\geq 2$ denotes higher-order nonlinear effective action, and $\kappa_p=(\kappa_{p,1},...,\kappa_{p,n})$ with $p=\{1,...,l\}$ represents different permutations of $l$ perturbing potentials appearing in the $n$th order perturbation theory.

The nonlinear $n$th order term of the effective action in the Matsubara representation is then,
\begin{align}
\mathcal{S}_{\kappa_p}^{(n\geq2)} =\frac{m_{\kappa_p}}{n}\sum_{k_1 \dots k_n}\tr\biggl[\biggl( \prod_{i=1}^{n-1}\mathcal{G}^0_{k_i} V^{\kappa_{p,i}}_{k_i-k_{i+1}} \biggr)\mathcal{G}^0_{k_n} V^{\kappa_{p,n}}_{{k_n-k_{i}}}\biggr],
\label{action_nth}
\end{align}
where $m_{\kappa_p}$ is the multiplicity of the permutation $\kappa_p$. In the above expression, we introduce the four-vector notation $k\equiv (\bm{k},i\omega_m)$, where $\bm{k}$ is the electron wave vector, $\omega_m=(2m+1)\pi/\beta$ is the $m$th fermionic Matsubara frequency, and $\mathcal{G}^0_{k}=(i \hbar \omega_m-\mathcal{H}_0)^{-1}$ is the noninteracting Green's function in the Matsubara representation.

\section{Model Hamiltonian and perturbing potentials}
We consider a generic unperturbed fermionic Hamiltonian for a 2D magnetic metal,
\begin{align}
\label{hamiltonian1}
\mathcal{H}_0=\mathcal{H}_{\mathrm{kin}}+\mathcal{H}_{\mathrm{so}}+\mathcal{H}_{\mathrm{sd}}, 
\end{align}
where $\mathcal{H}_{\mathrm{kin}}$ is the kinetic energy of itinerant electrons, $\mathcal{H}_{\mathrm{so}}$ is the SOC term, and the last term is the {\it{s-d}} exchange interaction between the spin of localized {\it{d}} electrons and itinerant {\it{s}} electrons. In the semiclassical regime, the {\it{s-d}} exchange interaction is modeled by
\begin{align}
\mathcal{H}_{\mathrm{sd}}[\bm{n}]=\mathcal{J}_{\mathrm{sd}} \bm{\Sigma} \cdot \bm{n}_{\bm{r},\tau}, \label{s-d exchange}
\end{align}
where $\mathcal{J}_{\mathrm{sd}}$ is the strength of the {\it{s-d}} exchange interaction, $\bm{\Sigma}$ is the spin matrix of itinerant electrons in an appropriate Hilbert space, and $\bm{n}$ is the spin order parameter field. In a simple FM metal, $\bm{\Sigma}\equiv{\bm{\sigma}}$ is the two-by-two spin Pauli matrix, and $\bm{n}$ is the magnetization direction vector, while in a simple two-sublattice AFM metal, $\bm{\Sigma}\equiv\bm{\sigma}\otimes\tau_z$, where $\tau_z$ denotes the sublattice degrees of freedom and $\bm{n}$ is the direction of the staggered N\'{e}el vector.

In our formalism for computing the DM interactions, the perturbing potential $V$ is a sum of {\it{i)}} the deviation of the order parameter from its equilibrium direction and {\it{ii)}} the ac electric field component of the laser pulse coupled to itinerant electrons.

{\it{i)}} The deviation of the order parameter from its equilibrium direction is defined as
\begin{align}
\delta\bm{n}_{\bm{r},\tau}=\bm{n}_0-\bm{n}_{\bm{r},\tau},
\end{align}
where $\bm{n}_0$ is the equilibrium direction of the order parameter.
Through the {\it{s-d}} exchange, the related perturbing potential is thus
\begin{align}
\label{pert1}
V^{(1)}=\mathcal{J}_{\mathrm{sd}} \bm{\Sigma} \cdot \delta\bm{n}_{\bm{r},\tau}\delta(\bm{r}-\bm{r}')\delta(\tau-\tau'). 
\end{align}

{\it{ii)}} The coupling between the electric field and the itinerant electrons is introduced in the noninteracting Hamiltonian via the minimal-coupling prescription, $\bm{p}\to\bm{p}+e\bm{A}_t$, where ${\bm{p}}=-i\hbar\bm{\nabla}$ is the linear momentum operator, $-e<0$ is the electron charge, and $\bm{A}$ is the vector potential related to the applied electric field $\bm{E}_t=-\partial_t\bm{A}_t$. The electric field perturbing potential in imaginary time \cite{Rammer} is
\begin{align}
\label{pert2}
V^{(2)}=\bm{j}_{\bm{r}}\cdot\bm{A}_\tau,
\end{align}
where $\bm{j}_{\bm{r}}= e \partial \mathcal{H}_{0}[\bm{n}_0]/\partial {\bm{p}}$ is the charge current density operator.

\section{Equilibrium DM interaction: Second-order perturbation theory}
In the absence of any external perturbation, SOC might induce an equilibrium DM interaction in the magnetic system. In this section, we find this equilibrium DM interaction. The results presented in this section have already been presented in Ref. \cite{PhysRevLett.120.197202}, but for completeness and further discussion, we outline them here briefly.

\begin{figure}
    \centering
  \includegraphics[width=\linewidth]{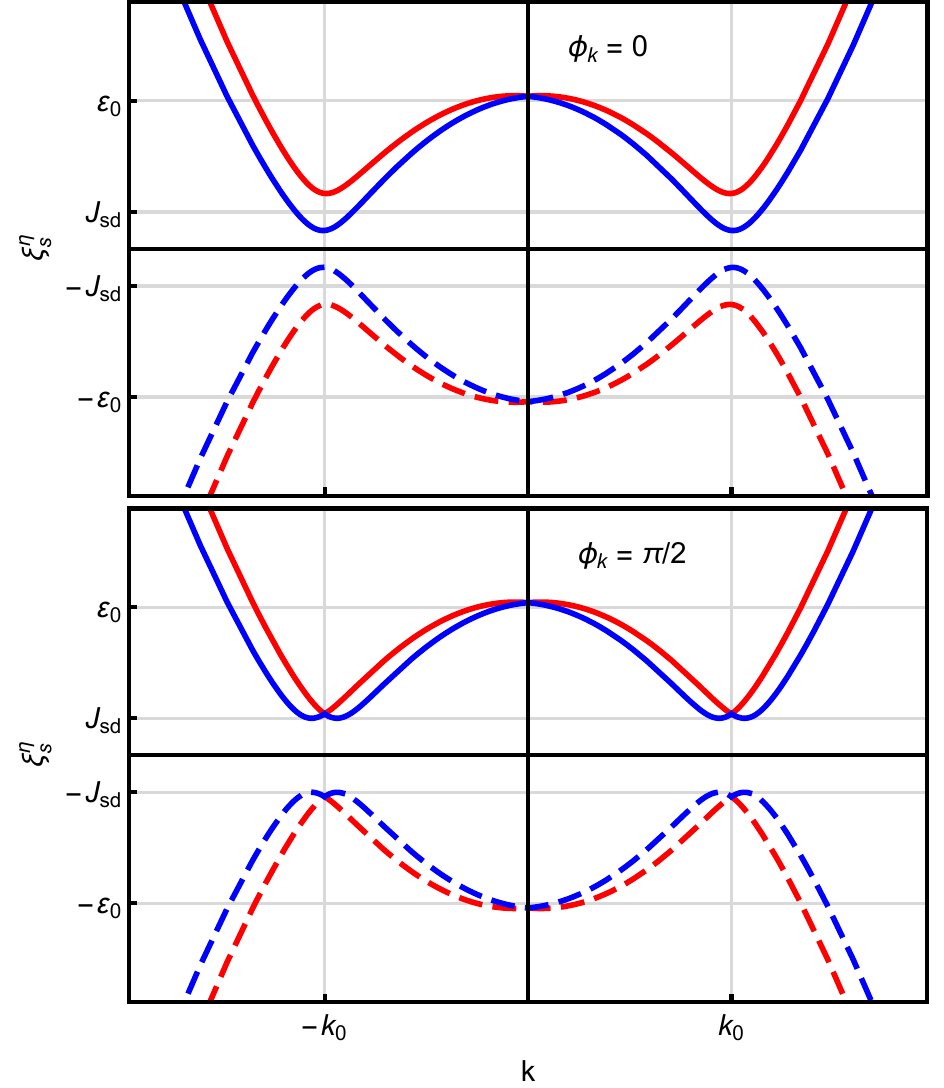}
\caption{The dispersion relation of a Rashba AFM metal with $\bm{n}_0=\hat{\bm{x}}$ in two different directions $\phi_k=0$ (top) and $\phi_k=\pi/2$ (bottom). The solid and dashed curves show the conduction ($\eta=+1$) and valence ($\eta=-1$) bands, respectively, while red and blue colors mark the $s=+1$ and $s=-1$ spin chirality of the subbands, respectively.}
  \label{fig:disp}
\end{figure}

Without loss of generality, we consider a 2D AFM metal with broken inversion symmetry in the $z$-direction normal to the plane modeled by a 2D Rashba SOC.
The effective unperturbed Hamiltonian of this system on a square lattice is \cite{PhysRevLett.120.197202}
\begin{align}
    \mathcal{H}_0 = \big(\gamma_{\bm{p}}\sigma_0 +i \alpha_\mathrm{R}\bm{\hat{z}}\cdot\boldsymbol{\sigma}\times\boldsymbol{\nabla}\big)\otimes\tau_x + \mathcal{J}_{\mathrm{sd}}\bm{\Sigma}\cdot\bm{n}_0,
    \label{hamiltonian}
\end{align}
where $\gamma_{\bm{p}}=-\varepsilon_0-\hbar^2{\bm{\nabla}}^2/2m$ is the kinetic energy operator, with $m$ denoting the effective electron mass, $\varepsilon_0=\hbar^2{\bm{k}}_0^2/2m$, $\alpha_\mathrm{R}$ is the Rashba SOC strength, $\bm{\Sigma}\equiv\bm{\sigma}\otimes\tau_z$, and $\bm{n}_0=\big(\sin\vartheta \cos\varphi, \sin\vartheta \sin\varphi,\cos\vartheta \big)$ is the equilibrium direction of the order parameter. The band dispersion of this system is anisotropic
$\xi^{\eta}_{s}=\eta \sqrt{\mathcal{J}^2_{\mathrm{sd}}+\gamma^2_{\bm{k}}+\alpha^2_\mathrm{R}k^2+s 2 \alpha_\mathrm{R} k \zeta_{\bm{k}}}$,
where
$\zeta_{\bm{k}}=\sqrt{\gamma^2_{\bm{k}}+\mathcal{J}^2_{\mathrm{sd}}(1-\sin^2{\vartheta}\sin^2{(\varphi-\phi_k)})}$, $k=|\bm{k}|$, $\phi_k=\tan^{-1}(k_y/k_x)$, $\eta=\pm1$ labels the conduction and valence bands, and $s=\pm1$ denotes the chirality of the subbands; see Fig. \ref{fig:disp}.

The perturbing potential [see Eq. (\ref{pert1})] is in the Fourier space given as
\begin{align}
\label{pert111}
V^{(1)}_{\bm{k},i\nu_l}=\mathcal{J}_{\mathrm{sd}} \bm{\Sigma} \cdot \delta\bm{n}_{\bm{k},i\nu_l}.
\end{align}

In the absence of any time-dependent perturbation, the lowest-order perturbation that leads to a finite DM interaction is the second-order perturbation theory. Using Eqs. (\ref{action5}) and (\ref{action_nth}), we find
\begin{align}
\mathcal{S}^{(2)}=\sum_{a,b}\sum_{\bm{q},m}\delta n^a_{\bm{q},i\nu_m}\Pi^{ab}_{\bm{q},i\nu_m}\delta n^b_{-\bm{q},-i\nu_m}, \label{action_eff}
\end{align}
with the following transverse spin susceptibility tensor,
\begin{align}
\label{Pi-ab}
\Pi^{ab}_{\bm{q},i\nu_m}=\frac{\mathcal{J}^2_\mathrm{sd}}{2}\sum_{\bm{k},n}\tr\left[\Sigma_a\mathcal{G}^0_{\bm{k},i\omega_n}\Sigma_b\mathcal{G}^0_{\bm{k}+\bm{q},i\omega_n+i\nu_m}\right],
\end{align}
where $\nu_m = 2m\pi/\beta$ is the $m$th bosonic Matsubara frequency.

The different elements of the DM tensor are calculated using Eq. (\ref{DMformula}). In the presence of an axially symmetric 2D Rashba SOC, the nonzero elements of the equilibrium DM tensor are only $D^{\rm{i}0}_x\equiv (D^0_{xxz}=-D^0_{zxx})$, $D^{\rm{i}0}_y\equiv (D^0_{yyz}=-D^0_{zyy})$, and thus, the DM interaction is {\it{interfacial}} type and is related to the dc spin susceptibility as \cite{PhysRevLett.120.197202}
\begin{align}
D_{x(y)}^{\rm{i}0} = i \frac{\partial \Pi^{x(y),z}_{\bm{q}=0}}{\partial q_{x(y)}}=-i \frac{\partial \Pi^{z,x(y)}_{\bm{q}=0}}{\partial q_{x(y)}}.
\label{dmi}
\end{align}
Since the assumed Rashba SOC, see Eq. (\ref{hamiltonian}), is isotropic, the obtained interfacial-type DM interaction is also isotropic, $D^{\mathrm{i}}_{0}\equiv (D^{\rm{i}0}_x=D^{\rm{i}0}_y)$. 
Finally, the micromagnetic DM free energy, Eq. (\ref{DMformula}), arisen from the isotropic Rashba SOC can be written in the form of Lifshitz invariants and in a compact form reads $E_{\mathrm{iDM}}=D^{\mathrm{i}}_{0} {\bm{n}}\cdot (\hat{\bm{z}}\times {\bm{\nabla}})\times {\bm{n}}$ \cite{Bogdanov}.

On the other hand, it can be shown that for a 2D Dresselhaus SOC, $\mathcal{H}_{\mathrm{SOC}}=i \alpha_\mathrm{D}\big(\sigma_x\partial_x-\sigma_y\partial_y\big)\otimes\tau_x$, where $\alpha_\mathrm{D}$ is the isotropic Dresselhaus SOC strength, the nonzero elements of the equilibrium DM tensor are $D^{{\rm{b}0}}_x\equiv (D^{0}_{zxy}=-D^{0}_{yxz})$, $D^{{\rm{b}0}}_y\equiv (D^{0}_{xyz}=-D^{0}_{zyx})$, 
with $D_{x}^{{\rm{b}0}} = i {\partial \Pi^{zy}_{\bm{q}=0}}/{\partial q_{x}}=-i {\partial \Pi^{yz}_{\bm{q}=0}}/{\partial q_{x}}$ and $D_{y}^{{\rm{b}0}} = i {\partial \Pi^{xz}_{\bm{q}=0}}/{\partial q_{y}}=-i {\partial \Pi^{zx}_{\bm{q}=0}}/{\partial q_{y}}$; thus, the DM interaction is {\it{bulk}} type. Consequently, the free energy of this isotropic bulk-type DM interaction can again be written in terms of Lifshitz invariants as $E_{\mathrm{bDM}}=D^{\rm{b}}_0 {\bm{n}}\cdot {\bm{\nabla}}\times {\bm{n}}$, with $D^{\rm{b}}_0\equiv (D^{\rm{b}0}_x=D^{\rm{b}0}_y)$ \cite{Bogdanov}. 

As we have already discussed, this DM interaction induced by SOC in metals has an RKKY-like origin \cite{Pyatakov_2014}. It arises from indirect anisotropic exchange interactions between two localized spins mediated by itinerant electrons.

Equations (\ref{Pi-ab}) and (\ref{dmi}) can in general be evaluated numerically for arbitrary band structures.
In the limit of small Rashba (Dresselhaus) SOC and at zero temperature, the equilibrium interfacial (bulk) DM interaction in the metallic and insulating regimes of an AFM system read \cite{PhysRevLett.120.197202},
\begin{equation}
\label{eq.DMI}
    D^{\mathrm{i (b)}}_0 = 
    \begin{cases}
    -\frac{k_0^2\mathcal{J}_{\mathrm{sd}}^2}{8\pi \epsilon_F^2}\big(2-\frac{\varepsilon^2_F}{\varepsilon_0^2}\big)\alpha_{\mathrm{R (D)}}, & \mathcal{J}_{\mathrm{sd}} < \varepsilon_F < \varepsilon_0\\
    \left(\frac{k_0^2}{2\pi}\right)\alpha_{\mathrm{R(D)}}, & 0 < \varepsilon_F < \mathcal{J}_{\mathrm{sd}}, 
\end{cases}
\end{equation}
where $\varepsilon_F$ is the Fermi energy.
These results show that the sign and amplitude of the DM interaction can be tuned by changing the Fermi level, e.g., via charge doping or a gate voltage \cite{PhysRevLett.120.197202}.

Note that as we have also found here, in equilibrium, it is forbidden to have a bulk-type (interfacial-type) DM interaction in the presence of a Rashba (Dresselhaus) SOC in our system. We later show that this argument is not anymore valid in the presence of the optical perturbation.

Here, we should emphasize that in the case of an arbitrary direction of the equilibrium order parameter $\bm{n}_0$ and in the presence of strong SOCs, there might be small anisotropy in the DM parameters along different spatial directions that we neglect in the present study \cite{PhysRevB.101.161403}.

\section{Light-induced DM interaction: Fourth-order perturbation theory}
In the presence of a laser pulse, the ac electric field component of the electromagnetic wave is coupled to itinerant charge carriers, and we can treat this coupling as a time-dependent perturbing potential in our formalism. To examine the effect of a laser pulse, we consider a spatially uniform time-oscillating electric field, described using a dipole approximation in the long-wavelength limit as,
\begin{align}
\bm{E}_\tau & = \mathcal{E}_\tau \bm{\hat\epsilon}, \nonumber\\
\mathcal{E}_\tau & = \mathcal{E}_0  e^{-\Omega \tau},
\end{align}
where $\mathcal{E}_0$ and $\Omega$ are the electric field amplitude and frequency of the laser pulse, respectively, and $\bm{\hat\epsilon}=(\epsilon_x \bm{\hat{x}}+\epsilon_y \bm{\hat{y}})/\sqrt{|\epsilon_x|^2+|\epsilon_y|^2}$ denotes the polarization direction with $\epsilon_{x(y)} \in \mathbb{C}$.
The perturbing potential is as follows, see Eq. (\ref{pert2}), 
\begin{align}
\label{pert22}
V^{(2)}&=\bm{j}_{\bm{r}}\cdot\bm{A}_\tau,\\
\bm{j}_{\bm{r}}&=-e\big(\frac{i\hbar\bm{\nabla}}{m}\sigma_0 + \alpha_\mathrm{R}\bm{\hat{z}} \times \bm{\sigma}\big) \otimes \tau_x.
\end{align}
In the frequency domain, we obtain
\begin{align}
\label{pert222}
V^{(2)}_{\bm{k},i\nu_l}&=u_{\bm{k}} \mathcal{E}_{-i\nu_l}+u^{\dagger}_{\bm{k}} \mathcal{E}_{i\nu_l},\\
u_{\bm{k}}&=\frac{\bm{j_k}\cdot\bm{\hat\epsilon}}{i\Omega}.
\end{align}

\begin{figure}[hbt!]
    \centering
\adjustbox{trim=0cm 0cm 2.0cm 0cm}{%
  \includegraphics[trim=4.2cm 0cm 0cm 0cm,clip,width=1.3\linewidth]{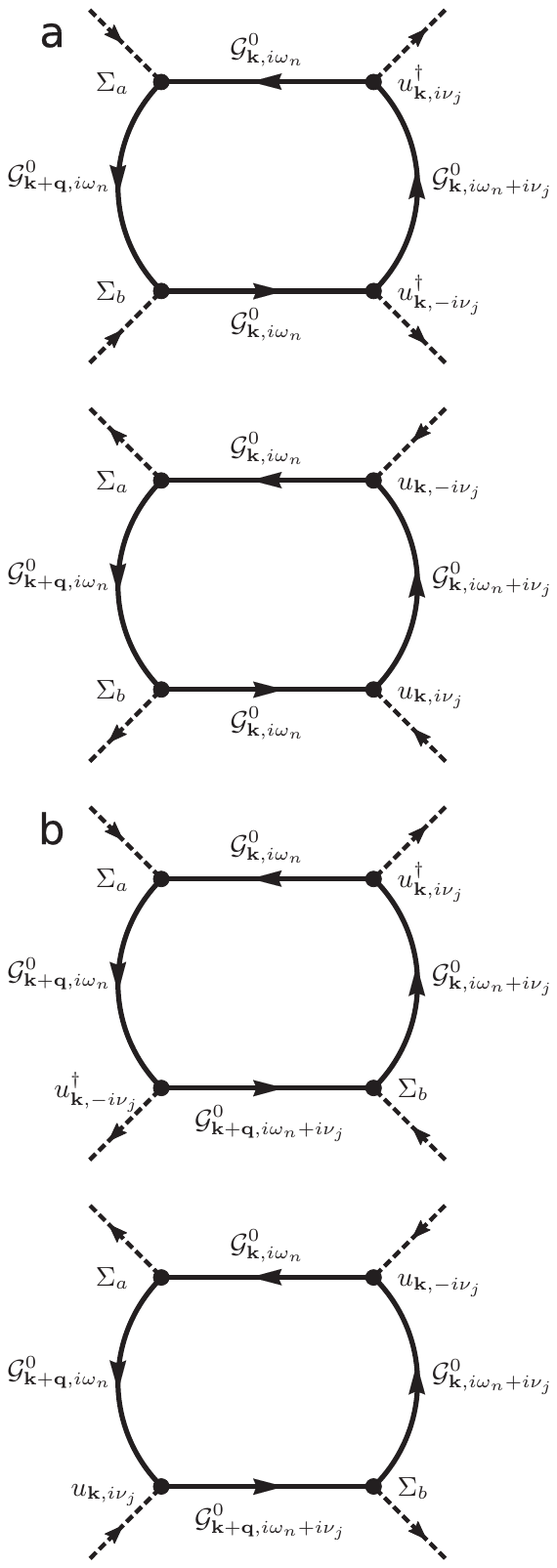}}
\caption{The Feynman diagrams of different photon-magnon scattering processes involved in the transverse correlation function computed within the fourth-order perturbation theory are shown, see Eq. (\ref{response-111}). (a) Successive and (b) overlapping processes in which two photons are created (top)/annihilated (bottom) and two magnons are annihilated (top)/created (bottom).
}
  \label{feynmann1}
\end{figure}

In the presence of an ac electric field, the lowest-order perturbation that leads to a finite static contribution to the equilibrium DM interaction is the fourth-order perturbation theory [see Eq. (\ref{action_nth})]
\begin{align}
\label{action4th}
\mathcal{S}_{\kappa_p}^{(4)} =& \frac{m_{\kappa_p}}{4}\sum_{k_1 \dots k_n}\tr\biggl[\biggl( \prod_{i=1}^{3}\mathcal{G}^0_{k_i} V^{{\kappa_{p,i}}}_{k_i-k_{i+1}} \biggr)\mathcal{G}^0_{k_n} V^{{\kappa_{p,n}}}_{{k_n-k_{1}}}\biggr]\\
=& \frac{m_{\kappa_p}}{4}\sum_{k_1, k_2, k_3, k_4}\tr\biggl[\mathcal{G}^0_{k_1} V^{\kappa_{p,1}}_{k_1-k_{2}} \mathcal{G}^0_{k_2} V^{\kappa_{p,2}}_{k_2-k_{3}}\mathcal{G}^0_{k_3}\nonumber\\&\times V^{\kappa_{p,3}}_{k_3-k_{4}}\mathcal{G}^0_{k_4} V^{\kappa_{p,4}}_{{k_4-k_{1}}}\biggr]\nonumber.
\end{align}
Since we have two perturbing potentials, $V^{(1)}$ and $V^{(2)}$ [see Eqs. (\ref{pert111}) and (\ref{pert222})], there are four different permutations $m_{\kappa_1}=4$ for $\kappa_1=(1,1,2,2)$ and two permutations $m_{\kappa_2}=2$ for $\kappa_2=(1,2,1,2)$.
Thus, the fourth-order action, using Eq. (\ref{action5}), can be expressed as
\begin{align}
\mathcal{S}^{(4)} =& \mathcal{S}_{\kappa_1}^{(4)}+\mathcal{S}_{\kappa_2}^{(4)},
\label{total-action_4th}
\end{align}
where
\begin{align}
\mathcal{S}_{\kappa_i}^{(4)} &= \hbar \beta \sum_{a,b}\sum_{{\bm{q}},l,m,j}
\delta n^a_{-\bm{q},-i\nu_l}\Pi^{i,ab}_{\bm{q},\nu_j,\nu_l,\nu_m}\delta n^b_{\bm{q},-i\nu_m},
\label{action_4th-1}
\end{align}
and the polarization tensors are given by,
\begin{align}
\Pi^{1,ab}_{\bm{q},\nu_j,\nu_l,\nu_m} = \frac{\mathcal{J}^2_{\mathrm{sd}}}{\hbar \beta}\sum_{{\bm{k}},n}
&\tr\biggl[\mathcal{G}^0_{\bm{k},i\omega_n} \Sigma_a \mathcal{G}^0_{\bm{k}+\bm{q},i\omega_n+i\nu_l} \nonumber\\ \times & \Sigma_b  \mathcal{G}^0_{\bm{k},i\omega_n+i\nu_l+i\nu_m} V^{(2)}_{\bm{k},-i\nu_j} \nonumber\\ \times & \mathcal{G}^0_{\bm{k},i\omega_n+i\nu_l+i\nu_m+i\nu_j} V^{(2)}_{\bm{k},i\nu_l+i\nu_m+i\nu_j}\biggr],\\
\Pi^{2,ab}_{\bm{q},\nu_j,\nu_l,\nu_m} = \frac{\mathcal{J}^2_{\mathrm{sd}}}{2\hbar \beta}\sum_{{\bm{k}},n}
&\tr\biggl[\mathcal{G}^0_{\bm{k},i\omega_n} \Sigma_a \mathcal{G}^0_{\bm{k}+\bm{q},i\omega_n+i\nu_l} \nonumber\\ \times & V^{(2)}_{\bm{k}+\bm{q},-i\nu_j}  \mathcal{G}^0_{\bm{k}+\bm{q},i\omega_n+i\nu_l+i\nu_j}  \Sigma_b \nonumber\\ \times & \mathcal{G}^0_{\bm{k},i\omega_n+i\nu_l+i\nu_m+i\nu_j} V^{(2)}_{\bm{k},i\nu_l+i\nu_m+i\nu_j}\biggr].
\label{response-1}
\end{align}
$\Pi^{1}$ is related to two successive electronic excitations and relaxations, while $\Pi^{2}$ involves overlapping excitation processes; see. Fig. \ref{feynmann1}.
Since the dynamics of the spins are much slower than those of electrons, we consider only quasistatic magnons, i.e., $\nu_l=\nu_m=0$. Using Eq. (\ref{pert222}), after some algebra, we can rewrite the total time-averaged quasistatic polarization tensor as
\begin{align}
\langle \Pi^{ab}_{\bm{q},\nu_j} \rangle &=\langle\Pi^{1,ab}_{\bm{q},\nu_j}\rangle + \langle\Pi^{2,ab}_{\bm{q},\nu_j}\rangle = \mathcal{J}^2_{\mathrm{sd}}\chi^{ab}_{\bm{q},\nu_j}\mathcal{E}_{i\nu_j}\mathcal{E}_{-i\nu_j}.
\label{response-11}
\end{align}
where $\langle ... \rangle$ denotes time averaging and $\chi$ is the ac response function tensor,
\begin{align}
\chi^{ab}_{\bm{q},\nu_j}&=\frac{1}{\hbar \beta}\sum_{{\bm{k}},n}
\tr\biggl[\mathcal{G}^0_{\bm{k},i\omega_n} \Sigma_a \mathcal{G}^0_{\bm{k}+\bm{q},i\omega_n} \Sigma_b  \mathcal{G}^0_{\bm{k},i\omega_n} u_{\bm{k}} \mathcal{G}^0_{\bm{k},i\omega_n+i\nu_j} u_{\bm{k}}
\nonumber\\ & +
\mathcal{G}^0_{\bm{k},i\omega_n} \Sigma_a \mathcal{G}^0_{\bm{k}+\bm{q},i\omega_n} \Sigma_b  \mathcal{G}^0_{\bm{k},i\omega_n} u^{\dagger}_{\bm{k}} \mathcal{G}^0_{\bm{k},i\omega_n+i\nu_j} u^{\dagger}_{\bm{k}}
\nonumber\\ & +\frac{1}{2}
\mathcal{G}^0_{\bm{k},i\omega_n} \Sigma_a \mathcal{G}^0_{\bm{k}+\bm{q},i\omega_n} u_{\bm{k}+\bm{q}}   \mathcal{G}^0_{\bm{k}+\bm{q},i\omega_n+i\nu_j} \Sigma_b \mathcal{G}^0_{\bm{k},i\omega_n+i\nu_j} u_{\bm{k}}
\nonumber\\ & +\frac{1}{2}
\mathcal{G}^0_{\bm{k},i\omega_n} \Sigma_a \mathcal{G}^0_{\bm{k}+\bm{q},i\omega_n} u^{\dagger}_{\bm{k}+\bm{q}}   \mathcal{G}^0_{\bm{k}+\bm{q},i\omega_n+i\nu_j} \Sigma_b \mathcal{G}^0_{\bm{k},i\omega_n+i\nu_j} u^{\dagger}_{\bm{k}}
\biggr].
\label{response-111}
\end{align}
In Fig. \ref{feynmann1}, the different Feynman diagrams related to this correlation function are plotted.

The contribution of the ac electric field to the interfacial-type DM interaction is given by the dynamical correlation functions through analytical continuation as,
\begin{align}
    \delta D^{\rm{i}}_{x(y)} = \left( \frac{\partial \chi^{x(y),z}_{\bm{q},\nu_j}|_{i\nu_j\to\Omega+i0^+}}{\partial q_{x(y)}}\right)_{\bm{q}\to0}\mathcal{J}^2_{\mathrm{sd}}\mathcal{E}^2_0.
\label{dmi_dynamic}
\end{align}
As we already discussed in the previous section, the bulk-type DM interaction $\delta D^{\rm{b}}_{x(y)}$ can also be computed using Eq. (\ref{dmi_dynamic}) through the proper change of the spatial indices.
Equations (\ref{response-11}), (\ref{response-111}) and (\ref{dmi_dynamic}) are our main results. 
These equations are quite general and might be implemented in first principle codes to compute induced DM interactions in materials with realistic band structures.

For a Rashba AFM model described by the Hamiltonian (\ref{hamiltonian}) and an AFM order parameter along the $x$ direction $\bm{n}_0=\hat{\bm{x}}$,
we  numerically and analytically compute the induced interfacial-type and possible bulk-type DM interactions. 
If the order parameter was perpendicular to the plane, i.e., along the direction of inversion symmetry breaking, we had an inplane rotational symmetry described by a $U(1)$ gauge symmetry. However, we consider a system with an order parameter along a particular direction inside the plane and thus, we reduce the rotational symmetry of the system further. In this case, we expect the response of the system to be sensitive to the laser polarization, and therefore the induced DM interaction might even be anisotropic depending on the laser pulse polarization. The ratio $\delta D_x/\delta D_y$ measures the DM anisotropy, and its value can be tuned by varying the Fermi level, SOC strength, and laser pulse frequency. Furthermore, in this far-from-equilibrium situation, it is now allowed to have a finite bulk-type (interfacial-type) DM interaction in a system with a Rashba (Dresselhaus) SOC.

For the light-induced interfacial-type DM interaction in the metallic regime and in the limits of high frequency ($\mathcal{J}_\mathrm{sd}\ll\varepsilon_F\ll\Omega$) and zero temperature, we find
\begin{equation}
\label{deltaDi}
    \delta D^{\rm{i}}_{x(y)} = 
    -\frac{\alpha_\mathrm{R} \mathcal{J}_\mathrm{sd}^2\left(\varepsilon_0^2-\varepsilon^2_F\right)}{2\pi\Omega^3\varepsilon_F^3}e^2 \lambda_{x(y)}\mathcal{E}_0^2,
\end{equation}
where $\lambda_{x(y)}={ {\rm{Re}}[3\epsilon_{x(y)}^2+\epsilon_{y(x)}^2]}/({|\epsilon_x|^2+|\epsilon_y|^2})$.

In the insulating regime ($\varepsilon_F<\mathcal{J}_\mathrm{sd}$) and in the nonresonant case $\Omega < 2\mathcal{J}_\mathrm{sd}$, we obtain,
\begin{align}
    \delta \tilde{D}^{\rm{i}}_{y} &= 
     \frac{2\alpha_\mathrm{R} \mathcal{J}^2_\mathrm{sd}}{\pi\Omega^3\sqrt{4\mathcal{J}^2_\mathrm{sd}-\Omega^2}}\tan^{-1}{\left(\frac{\Omega}{\sqrt{4\mathcal{J}^2_\mathrm{sd}-\Omega^2}}\right)}e^2 \tilde{\lambda}\mathcal{E}_0^2,\\
       \delta \tilde{D}^{\rm{i}}_{x} &=  \frac{2\mathcal{J}^2_\mathrm{sd}-\Omega^2}{4\mathcal{J}^2_\mathrm{sd}-\Omega^2} \delta \tilde{D}^{\rm{i}}_{y}-\frac{\alpha_\mathrm{R} \mathcal{J}^2_\mathrm{sd}}{\pi\Omega^2 \left(4 \mathcal{J}^2_\mathrm{sd}-\Omega ^2\right)}e^2 \tilde{\lambda}\mathcal{E}_0^2,
\end{align}
where $\tilde{\lambda}={ {\rm{Re}}[\epsilon_{x}^2+\epsilon_{y}^2]}/({|\epsilon_x|^2+|\epsilon_y|^2})$. 

These results show that a nonresonant polarized laser pulse can induce an anisotropic DM interaction. In the insulating phase, the induced DM interaction is zero for circularly polarized laser pulses in our model.

\begin{figure}
    \centering
  \includegraphics[width=\linewidth]{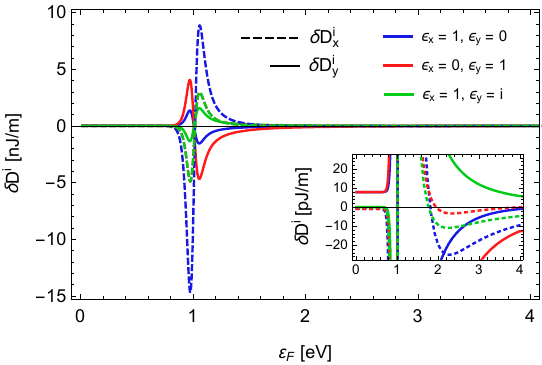}
\caption{Numerical results for the laser-induced interfacial-type DM interaction in a Rashba AFM system as a function of the Fermi energy $\varepsilon_F$, for a laser pulse with linear polarization along $x$ (blue) and $y$ (red) directions, and circular polarization (green). The two components of the DM tensor related to the interfacial DM interaction are plotted by dashed lines $\delta D_{\rm{x}}^{\rm{i}}$ and solid lines $\delta D_{\rm{y}}^{\rm{i}}$.
The equilibrium direction of the AFM order parameter is $\bm{n}_0=\hat{\bm{x}}$. The inset shows the induced DM interaction inside and far above the gap. For the numerical calculations, we use $\varepsilon_0=4\,\mathrm{eV}$, $\mathcal{J}_\mathrm{sd}=1\,\mathrm{eV}$, $k_0\alpha_R = 0.4\,\mathrm{eV}$, $a=0.4\,\mathrm{nm}$, $\Omega=0.75\,\mathrm{eV}$, and $T=300\,\mathrm{K}$.}
  \label{fig:num_y}
\end{figure}

\begin{figure}
    \centering
  \includegraphics[width=\linewidth]{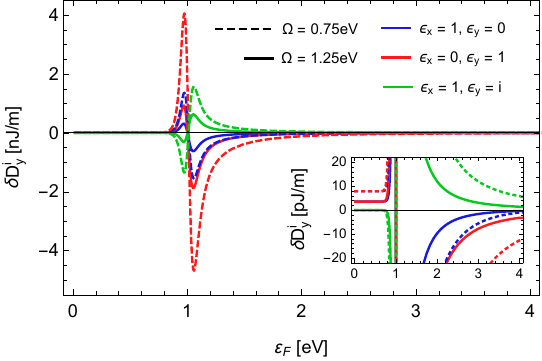}
\caption{The frequency dependence of the laser-induced interfacial-type DM interaction in a Rashba AFM system as a function of the Fermi energy $\varepsilon_F$. The induced DM interaction changes sign and peaks in magnitude at the insulator-metal transition. The system parameters are the same as in Fig. \ref{fig:num_y}.}
  \label{fig:num}
\end{figure}

The numerical results for different laser pulse polarizations and frequencies at room temperature are shown in Figs. \ref{fig:num_y} and \ref{fig:num}. These results show that not only can we change the amplitude and sign of the induced DM interactions, but we can also tune the DM anisotropy in this system.

\begin{figure}
    \centering
  \includegraphics[width=\linewidth]{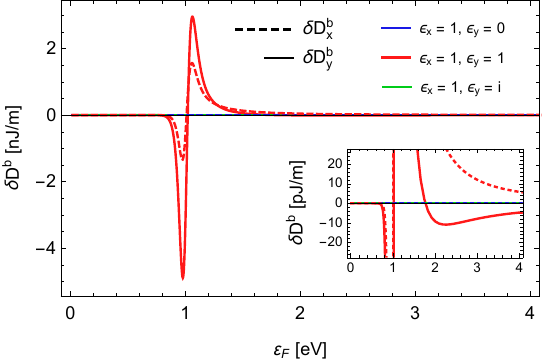}
\caption{Numerical results for the laser-induced bulk-type DM interaction in a Rashba AFM system as a function of the Fermi energy $\varepsilon_F$ for different laser polarizations. The system parameters are the same as in Fig. \ref{fig:num_y}.}
  \label{fig:num_b}
\end{figure}

The Rashba SOC in equilibrium gives only rise to an interfacial-type DM interaction and thus, a bulk-type DM interaction is forbidden, see Sec. V. However, in the nonequilibrium situation other elements of the DM tensor, see Eqs. (\ref{response-111}) and (\ref{dmi_dynamic}), can be nonzero depending on the order parameter direction and the laser polarization. 
In the metallic regime, up to linear order in the Rashba SOC strength, we find
\begin{equation}
    \delta D^{\rm{b}}_{x(y)} = 
    -\frac{\alpha_\mathrm{R} \mathcal{J}_\mathrm{sd}^2\left(\varepsilon_0^2-\varepsilon^2_F\right)}{\pi\Omega^3\varepsilon_F^3}e^2 \lambda^{\rm{b}}_{x(y)}\mathcal{E}_0^2,
\end{equation}
where $\lambda^{\rm{b}}_{x(y)}={ {\rm{Re}}[\epsilon_{x(y)}\epsilon_{y(x)}]}/({|\epsilon_x|^2+|\epsilon_y|^2})$.
This result shows that in a Rashba AFM system, a bulk-type DM interaction might be induced for certain laser pulse polarizations. However, the bulk-type DM interaction is zero inside the gap in our model. These analytical results are also confirmed with numerical calculations, see Fig. \ref{fig:num_b}.

\section{Summary and concluding remarks}
In this study, we investigated the effect of nonresonant laser pulses on the DM interactions. The ultrafast renormalization of DM interactions can destabilize the magnetic ground state and excite magnons in the systems. Additionally, a sudden change in DM interactions in magnetic systems with topological solitons, stabilized with DM interactions, may trigger ultrafast dynamics of these chiral objects. On the other hand,  generation of static DM interactions of different symmetries, bulk-type and interfacial-type, in the far from equilibrium magnetic systems may lead to dynamical stabilization of novel exotic topological spin textures.

In the present calculations, we ignored the effect of laser-induced heating. To avoid too much sample heating, it is possible to use nonresonant {\it{subpicosecond}} laser pulses with {\it{high}} frequencies. In the presence of heating, we have to take into account the effect of thermal magnons in our formalism as well. On the other hand, in quantum AFM systems with weak or zero magnetic anisotropy, quantum fluctuations are pronounced at zero temperature. In this study, the quantum fluctuations have been neglected since we considered an AFM system with strong anisotropy that makes the system assume a classical N{\'e}el ground state.
An advantage of our formalism is that both quantum and thermal effects can be implemented in this machinery method which will be the focus of our future work.

Additionally, we should emphasize that although in the current study, we considered only the direct coupling of photons and spin-spin interactions by the renormalization of the DM parameters, there is another nonthermal optomagnetic mechanism known as the inverse Faraday effect that {\it{indirectly}} couples photons and spins via SOC by inducing an effective magnetic field along the laser pulse propagation \cite{Kalashnikova_2015, RevModPhys.82.2731,metallicAFM}. This effective magnetic field couples to spins by a Zeeman-like interaction and excites magnons. Both mechanisms are second order in the electric field, and the competition between these two mechanisms is an open question for further studies.

In summary, we developed a theory to compute the light-induced DM interactions induced by an ac electric field in AFM systems. The formalism is quite general and may be used in first-principle codes for complicated band structures, and to compute DM interactions in novel states of magnetic materials such as topological AFM materials and Weyl AFM systems.
Using a Rashba AFM model, we evaluated the amplitude of the light-induced DM interaction. Our theory shows that ultrafast laser pulses can suddenly change the equilibrium DM interaction and thus lead to ultrafast magnon excitation in AFM systems. We also showed that in nonequilibrium situations, it is possible to have different types of DM interactions which are usually forbidden in equilibrium for a particular system. The results suggest that nonequilibrium magnetic systems may host interesting and exotic physics which are promising for future spintronics.

\section*{Acknowledgement}
This work was supported by the Norwegian Financial Mechanism 2014-2021 Project No. 2019/34/H/ST3/00515, ``2Dtronics'', and the Research Council of Norway through its Centres of Excellence funding scheme, Project No. 262633, ``QuSpin''.

\bibliography{bi}

\end{document}